\documentclass{aa}
\usepackage[varg]{txfonts}
\usepackage{graphicx}
\usepackage{txfonts}
\usepackage{natbib}
\usepackage{verbatim}
\usepackage[pdftex]{color}
\bibpunct{(}{)}{;}{a}{}{,} 

\usepackage{booktabs}
\definecolor{ultramarine}{rgb}{0.07, 0.1, 0.6} 
\definecolor{myblue}{rgb}{0.07, 0.2, 0.6} 
\definecolor{dopal}{rgb}{.70, .25, .05}

\begin{document}
%
\title{The impact of Coulomb diffusion of ions on the pulsational properties of DA 
white dwarfs}


\author{Leandro G. Althaus \inst{1,2},
  Alejandro H. C\'orsico \inst{1,2}, and
Francisco De Ger\'onimo\inst{1,2}        }
\institute{Grupo de Evoluci\'on Estelar y Pulsaciones. 
           Facultad de Ciencias Astron\'omicas y Geof\'{\i}sicas, 
           Universidad Nacional de La Plata, 
           Paseo del Bosque s/n, 1900 
           La Plata, 
           Argentina
           \and
           CCT - CONICET
           }
\date{Received ; accepted }

\abstract{Element diffusion is a key physical process that substantially impacts the superficial abundances, 
internal structure, pulsation properties, and  evolution of white dwarfs.}
{We study the effect of Coulomb separation of ions in the cooling times of evolving white dwarfs, 
their chemical profiles, the Brunt-V\"ais\"al\"a (buoyancy) frequency, and the pulsational 
periods at the  ZZ Ceti instability strip. }
{We follow the full evolution of white-dwarf models in the range $0.5-1.3\ M_\sun$ derived from 
their progenitor history on the basis of a time-dependent  element diffusion scheme that 
incorporates the effect of gravitational settling of ions due to Coulomb interactions at 
high densities. We compare the results for the evolution and pulsation periods of ZZ Ceti 
stars with the case where this effect is neglected.}
{We find that Coulomb sedimentation profoundly alters the chemical profiles of ultra-massive 
($M_{\star}\gtrsim1 M_\sun$) white dwarfs along their evolution, preventing 
helium from diffusing inward toward the core, and thus leading to much narrower 
chemical transition zones. As a result, significant changes in the $g$-mode pulsation periods as 
high as $15 \%$ are expected for ultra-massive ZZ Ceti stars. For less-massive white dwarfs, 
the impact of Coulomb separation is much less noticeable, inflicting period changes in ZZ Ceti 
stars that are below the period changes that result from uncertainties in progenitor
evolution, albeit larger than typical uncertainties of observed periods.}
{Coulomb diffusion of ions profoundly affects the diffusion flux in ultra-massive white dwarfs, 
driving  the  gravitational settling of ions with the same $A/Z$ (mass to charge number). 
We show that it strongly alters the period spectrum of such white dwarfs, which should   
be taken into account in detailed asteroseismological analyses of ultra-massive ZZ Ceti stars.}

\titlerunning{The impact of Coulomb diffusion on the pulsational properties of white dwarfs}
\authorrunning{Althaus et al.}

\maketitle

\section{Introduction}
\label{introduction}

Element diffusion  is a  key physical  process that  may substantially
modify  the  superficial  abundances,  internal chemical structure,  pulsation
properties, and  evolution of  a large variety  of stars  at different
evolutionary      stages,       including      our       Sun 
\citep[see][]{2015ads..book.....M}.    Because   of  their   extremely   
large gravities, this  is particularly true in  the case of white-dwarf 
(WD) stars, for  which diffusion driven  by gravity is responsible  for the
purity  of the  outer layers  that  characterizes most  of these  stars
\citep[see][for                          reviews]{2008ARA&A..46..157W,
  2010A&ARv..18..471A,2019A&ARv..27....7C}.   Also,  at  the  chemical
interfaces of these stars,  characterized by large composition gradients,
chemical diffusion tends to dominate over diffusion driven by gravity,
thus strongly smoothing out the chemical profiles during WD evolution,
which have  consequences for  the predicted pulsational  properties of
these  stars.  Earlier  studies  by the La  Plata  Group demonstrated  the
necessity of a proper treatment of element diffusion for an assessment
of  the  adiabatic oscillation properties of pulsating WDs and the  mode-trapping  features  produced  by the  outer  chemical transition  regions \citep[see][]{2002A&A...387..531C,     2004A&A...417.1115A}.     
Also, diffusion   processes  are   at  the   root  of   the  occurrence   of
diffusion-induced novae, i.e., WDs that under certain  circumstances experience
 a  thermonuclear  flash  induced by  chemical  diffusion  that
carries    hydrogen (H)   to    deeper    and    hotter    layers
\citep[][]{2011MNRAS.415.1396M}. Finally,  gravitational settling  of
minor species such as $^{22}$Ne in  the carbon/oxygen (CO) core of WDs
releases enough energy so as to impact substantially the cooling times
of     WDs  \citep{2002ApJ...580.1077D, 2010Natur.465..194G}.

\cite{2010ApJ...723..719C} and \cite{2013PhRvL.111p1101B} have studied
the  diffusion  process  in  strongly-coupled  Coulomb  plasmas.   In
particular,  \cite{2013PhRvL.111p1101B}  have explored  the  diffusion
currents  for  strongly-coupled  Coulomb  mixtures  of ions  in  dense
stellar matter typical of compact stars such as WDs and neutron stars.
These  authors extended  the work  of \cite{2010ApJ...723..719C}  to the
case  of  non-equilibrium mixtures,  and  showed  that Coulomb  separation
affects  the  diffusion  flux  in WDs,  driving  the  gravitational
settling of ions with the same $A/Z$ (mass to charge number). Indeed, in
such mixtures, the contribution of  gravity and induced electric field
to sedimentation is  negligible, and the effect  of Coulomb separation
becomes relevant.   They find that ions with larger $Z$
move  to  deeper layers,  being  this  effect  stronger for  a  larger
difference  of $Z$  in the  mixture  and for  larger gravities,  i.e.,
massive  WDs. In  a more  recent paper, \cite{2020A&A...635A.103K}  
have  implemented the Coulomb corrections following  \cite{2013PhRvL.111p1101B} 
to study the problem of carbon dredge-up in helium (He)-rich WDs.

\cite{2013PhRvL.111p1101B} suggest that the redistribution of ions due
to Coulomb  separation could affect  the thermal evolution of  WDs and
their  pulsational  properties,  thus  impacting  the  predictions  of
asteroseismology. Our paper  is aimed at precisely  studying the impact
of Coulomb separation of ions on  the evolution and pulsations of
ZZ  Ceti   stars.  These   pulsating  stars  are   H-rich (DA spectroscopic 
class)  WDs with $10\,400$ K  $\lesssim T_{\rm  eff} 
\lesssim  12\,400$ K  and  $7.5  
\lesssim \log  g  \lesssim 9.1$, characterized   by   multi-periodic   
luminosity  variations   due   to $g$(gravity)-mode  pulsations 
\citep[see][for a recent review]{2019A&ARv..27....7C}.   
In  particular, ultra-massive ZZ Ceti stars ---where Coulomb separation of ions
would be more  relevant and the impact on their pulsation periods could be
larger--- have  been    reported
\citep{2005A&A...432..219K,2010MNRAS.405.2561C,2013MNRAS.430...50C,
  2013ApJ...771L...2H,2017MNRAS.468..239C,2019MNRAS.486.4574R}.
Specifically, we  study here the  evolution of WD  models with 
masses in the  range $0.5-1.3\ M_\sun$ derived
from their progenitor  history. The WD mass values cover
the range of  relevant stellar masses expected for these stars.  We consider a
time-dependent   element  diffusion   scheme  that   incorporates  the
gravitational  settling  term  due  to Coulomb  interactions  at  high
densities.  We assess the impact  of Coulomb separation on the cooling
times of our  WD models and on their  chemical profiles, Brunt-V\"ais\"al\"a
frequency, and pulsational periods.  The paper is organized as follows.
In  Sect.~\ref{codes} we  describe the  input physics  of our  stellar
models, in  particular the  modification to  our treatment  of element
diffusion to include  the Coulomb interaction effect.  In that section
we also  discuss the impact  of Coulomb sedimentation on  the evolving
chemical  profiles.  In  Sect.~\ref{pulsation_results}  we
describe the impact of Coulomb diffusion on the pulsational properties
of DA WDs.  Finally,  in Sect.~\ref{conclusions}  we summarize  the main
findings of the paper.

\section{Input physics and white dwarf models}
\label{codes}

\subsection{Numerical codes}
\label{num_codes}

The pulsational properties of the WD models presented in this work are based
on the evolutionary models  provided by the {\tt LPCODE}  stellar evolution code.
This code has  been amply used and tested in numerous  stellar evolution contexts of 
low-mass and, particularly WD stars \citep[see][for         details]
{2003A&A...404..593A,2005A&A...435..631A, 2013A&A...555A..96S, 2015A&A...576A...9A,
2016A&A...588A..25M,2020A&A...635A.164S,
2020A&A...635A.165C}. Specifically,  {\tt  LPCODE} considers 
a full  treatment of  energy  sources, including the  energy
contribution due to phase  separation of core  chemical species
upon crystallization. The treatment  of crystallization is  based on
the phase diagrams of  \cite{2010PhRvL.104w1101H} for
dense  CO mixtures,  and  that of  \cite{2010PhRvE..81c6107M} for  ONe
mixtures. Relevant for this work,  {\tt  LPCODE} computes WD evolution
in a self-consistent way with the changes in the  internal chemical distribution
that result from the mixing of  all the core chemical components induced by
the  mean molecular  weight  inversion, element  diffusion (see below), and  phase
separation of  core chemical  constituents upon  crystallization (and the ensuing
mixing in the layers above the crystallized core). Concerning this work, abundance
changes during crystallization  is a relevant
issue in the case of pulsating ultra-massive ZZ Ceti stars that are expected to be mostly crystallized.
In  {\tt  LPCODE}, these abundance changes are assessed in a self-consistent way with
WD evolution. Energy release resulting from $^{22}$Ne sedimentation
was not considered in this study.

For  the pulsational  analysis we use  the {\tt  LP-PUL} pulsation
code described in \citet{2006A&A...454..863C}. This code  employs the "hard-sphere" boundary
conditions to  account for the  effects   of  crystallization   on  the  pulsation   spectrum  of
$g$ modes.  These   conditions  assume  that  the   amplitude  of  the
radial and horizontal eigenfunctions  of  $g$-modes   is  null  below the  solid/liquid
interface because of  the non-shear modulus of the  solid, as compared
with  the fluid  region \citep[see][]{1999ApJ...526..976M}.  The central boundary 
conditions are located at the mesh-point  of  the crystallization front
\citep[see][]{2004A&A...427..923C,2005A&A...429..277C,2019A&A...621A.100D,2019A&A...632A.119C}.
The    Brunt-V\"ais\"al\"a    frequency     is    computed    as    in
\cite{1990ApJS...72..335T}.  The  computation of  the Ledoux  term $B$
includes the effects of having  multiple chemical species that vary in
abundance.

\subsection{Diffusion treatment in a strongly coupled Coulomb plasma of ions}
\label{coulom}

{\tt LPCODE} considers a new full-implicit treatment of time-dependent
element  diffusion that  includes thermal  and chemical  diffusion and
gravitational settling \citep{2020A&A...633A..20A}. In this work,
the follow the diffusion for the isotopes  $^1$H,$^3$He, $^4$He, $^{12}$C,  $^{13}$C,
$^{14}$N,$^{15}$N, $^{16}$O, $^{17}$O,$^{18}$O, $^{19}$F,  $^{20}$Ne, $^{22}$Ne, $^{23}$Na,
and $^{24}$Mg. Our treatment of
diffusion  is based  on  the  formalism of  \cite{1969fecg.book.....B}
that  provides the  diffusion velocities  in a  multi-component plasma
under the influence of gravity, partial pressure, and induced electric
fields.  Here,   diffusion  velocities   satisfy  the  set   of  $N-1$
independent equations for ions\footnote{For  the sake of simplicity,  we do not
write here the equations that describe thermal diffusion.}:

\begin{equation}
\frac{dp_i}{dr}-\frac{\varrho_i}{\varrho}\frac{dP}{dr}-n_iZ_ieE=
\sum_{j\ne i}^NK_{ij}\left(w_j-w_i\right),
\label{diff1}
\end{equation}

\noindent where  $p_i$, $\varrho_i$, $n_i$, $Z_i$, and $w_i$ denote,
respectively, the ion partial pressure, mass density, number density, mean
charge,   and  diffusion velocity for  chemical species $i$. $N$  is the
number  of ionic  species plus  electrons,  $e$ the charge unit and $E$ the electric
field. Resistance coefficients $K_{ij}$ are
from  \cite{1986ApJS...61..177P}.  This  set of  equations  is  solved
together with the  equations for no net  mass flow $\sum_iA_in_iw_i=0$,
with $A_i$ being the atomic mass number, and no electrical current $\sum_iZ_in_iw_i=0$.

Using  $dP/dr=-g \varrho$ and  $\varrho_i= A_i n_i m_{\rm u}$, Eq.~(\ref{diff1})
can be written as 

\begin{equation}
\frac{1}{n_i}\sum_{j\ne i}^NK_{ij}\left(w_i-w_j\right) - Z_ieE=
 -A_i m_{\rm u} g -\frac{1}{n_i} \frac{dp_i}{dr},
 \label{difff}
\end{equation}

\noindent where $g$ is  the  gravitational acceleration and  $m_{\rm u}$ the atomic mass. Diffusion velocities at each depth of the star are found as described in \cite{2020A&A...633A..20A}. To include the Coulomb interaction effect
we proceed as in \cite{2020A&A...635A.103K} and \cite{2013PhRvL.111p1101B}, and write 

\begin{equation}
\frac{dp_i}{dr}= \frac{dp^{\rm ideal}_i}{dr} + n_i \frac{d\mu^{\rm coul}_i}{dr},
\end{equation}

\noindent where $p^{\rm ideal}_i$ and  $\mu^{\rm coul}_i$ are
the ideal  pressure and the
chemical potential due to Coulomb interactions of ion $i$.  $\mu^{\rm coul}_i$ is  given by
\cite{2013PhRvL.111p1101B} as

\begin{equation}
  \mu^{\rm coul}_i= -0.9 \frac{Z_i^{5/3} e^2}{a_e}, 
\end{equation}

\noindent with $a_e= (4 \pi n_e/3)^{-1/3}$ being the electron-sphere radius.  Hence,

\begin{equation}
  \frac{dp_i}{dr}= \frac{dp^{\rm ideal}_i}{dr} - n_i\ 0.3 \frac{Z_i^{5/3} e^2}{a_e}
  \frac{1}{n_e}\frac{dn_e}{dr}.
\end{equation}

This  treatment is  justified in  the regime  of strong  ion coupling.
Coulomb sedimentation may be relevant in a mixture of ions with  the same $A/Z$
(such as  $^{4}$He, $^{12}$C, and  $^{16}$O).   We can
illustrate this  by simplifying  the set  of Eq.~(\ref{difff})  to the
case of two  ion species ($i=1,2$). Assuming  no temperature gradients
and electron  to have zero  mass, the diffusion  flux of ion  2 ($J_2=
\varrho_2w_2$)   can   be   expressed  as   the   contributions   from
gravitational settling, Coulomb diffusion,  and ordinary diffusion due
to chemical gradients as

\begin{equation}
  J_2= J_2^{\rm grav} + J_2^{\rm coul} + J_2^{\rm diff},
   \label{flux}
\end{equation}

\noindent where

\begin{equation}
  J_2^{\rm grav} =  \varrho_2\  \frac{\varrho_1 n Z_1Z_2D}{\varrho n_e K_BT}\left({\frac{A_1}{Z_1}} -
  {\frac{A_2}{Z_2}} \right)m_{\rm u} g,
  \label{grav}
\end{equation}

\begin{equation}
  J_2^{\rm coul} =  \varrho_2\  \frac{\varrho_1 n D Z_1Z_2}{ n_e K_BT}g \left(Z_1^{2/3} - Z_2^{2/3}\right) 
  \frac{0.3 e^2}{a_e P \gamma},
    \label{coul}
\end{equation}

\noindent and 

\begin{equation}
  J_2^{\rm diff} =  \frac{D m_1 m_2 n}{ n_e\varrho} \left( n_2Z_2\frac{dn_1}{dr} - n_1Z_1 \frac{dn_2}{dr}\right).
    \label{diff}
\end{equation}

Here,  $D$   is   the   standard  diffusion   coefficient,
$m_i$=$A_im_u$, and $\gamma={\rm \partial\,  ln}\, P / {\rm \partial\,
  ln }\,  \varrho$\footnote{Our expression  for the diffusive  flux is
  similar      to      that       shown      in      Eqs.17-19      of
  \cite{2013PhRvL.111p1101B}.}.  Pressure  has   been  assumed  to  be
provided  by   strongly  degenerate   electrons.  We note   that  Coulomb
interaction contributes  to the gravitational  settling in such  a way
that ions with  larger $Z$ move to deeper layers.   We note also that for
plasmas made of  a mixture of $^{4}$He and  $^{12}$C, $J_2^{\rm grav}$
vanishes, and  Coulomb separation  should be  important in
driving gravitational settling of ions.  This is particularly true
in the case of more massive (and compact) WDs characterized by
larger gravities \citep{2013PhRvL.111p1101B}. In fact, we will see that Coulomb sedimentation
cannot be neglected in ultra-massive WDs. In  the case of a WD envelope
made of $^{1}$H  and $^{4}$He, $J_2^{\rm grav}$  dominates and Coulomb
separation constitutes a minor contribution to the settling of ions.

\subsection{White dwarf models}
\label{wd}

To assess  the impact of  Coulomb diffusion,  we have computed  the WD
evolution considering and disregarding the Coulomb interaction term in
the  diffusion equations.   We consider  four WD  models extracted from
the evolution of single progenitors with final   stellar
masses of  0.576, 0.833, 1.159 and  1.292\,$M_{\sun}$.  The starting
0.576 and 0.833 $M_{\sun}$  WD configurations, characterized by CO
cores,  are   H-burning  post asymptotic giant branch (AGB)  star  models   computed  by
\cite{2016A&A...588A..25M} from the full evolution of their progenitor
stars  with metallicity  $Z$=0.02  and  initial masses  of  
1.5 and  4.0 $M_{\sun}$,  respectively.   Such  starting  WD models  are  based  on
detailed and  improved micro-  and macrophysics processes  involved in
AGB  and thermally-pulsing AGB  modeling,  of relevance  for  the predicted  internal
composition  of   WDs.  The   starting  1.159   and  1.292\,$M_{\sun}$
ultra-massive  WDs were  taken from  \cite{2019A&A...625A..87C}. These
WDs  result  from  off-center
carbon burning in the  single-evolution scenario.  Their
progenitors experience a violent carbon-ignition phase followed by the
development of an inward-propagating  convective flame that transforms
the  CO core  into a  degenerate  ONe one.  Such WDs  are composed  of
$^{16}$O and  $^{20}$Ne ---with  traces of $^{23}$Na  and $^{24}$Mg---
\citep[see][]{2007A&A...476..893S,2010A&A...512A..10S}.   All   of  the
WD models  were evolved  from the  beginning of  cooling track,  where we
performed  the  core mixing  implied  by  the  inversion of  the  mean
molecular  weight  left by  prior  evolution,  down to  low  effective
temperatures, in  a consistent  way with abundances  changes resulting
from   element  diffusion,   phase   separation  on   crystallization,
convective mixing and  residual nuclear  burning in the case of CO
WD models.

\begin{figure}
        \centering
        \includegraphics[width=1.\columnwidth]{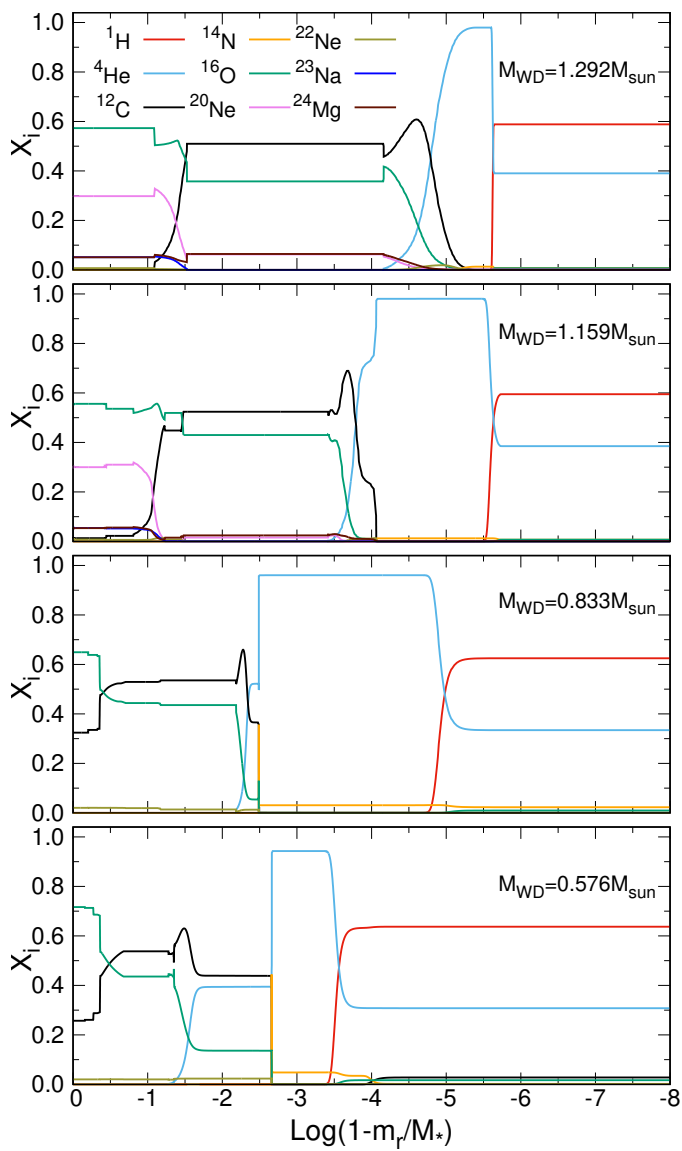}
        \caption{Abundance by mass of $^1$H, $^4$He, $^{12}$C, $^{14}$N,
          $^{16}$O, $^{20}$Ne, $^{22}$Ne, $^{23}$Na, and $^{24}$Mg
          versus the outer mass coordinate  for our  0.576, 0.833,
          1.159 and  1.292\,$M_{\sun}$ WD models at the beginning of
          the cooling track.} 
        \label{perfiles_inicial.eps}
\end{figure}

The  chemical profiles  of our  initial WD  models are  shown in  Fig.
\ref{perfiles_inicial.eps}  for some  selected isotopes.  The internal
composition of  WDs is a crucial  aspect for the determination  of the
pulsational  properties of  these stars.  The WD  models shown  in the
figure correspond at the beginning of their cooling phase prior to the
onset  of element  diffusion  and after  the core  mixing  due to  the
inversion of the mean molecular  weight. Their core chemical structure
reflects  the  nuclear burning  and  mixing  history along  progenitor
evolution. We note that  both the H and He contents  of the WDs
decrease with the stellar mass\footnote{For the 1.292\,$M_{\sun}$ WD
  model,  the  H  envelope was  artificially  imposed,  without
  considering  nuclear  burning  processes.}.   Also is noted  the  large
abundance of $^{14}$N  in the He-rich buffer of the  CO WD models,
in  particular for  the 0.576  \,$M_{\sun}$ WD  model, reflecting  the
occurrence of  appreciable third dredge-up.  This  model experiences a
final thermal  pulse at  the very  end of the  thermally pulsing AGB phase that is
responsible    for    larger    surface   carbon    abundances
\citep[see][]{2016A&A...588A..25M}.

\begin{figure}
        \centering
        \includegraphics[width=1.0\columnwidth]{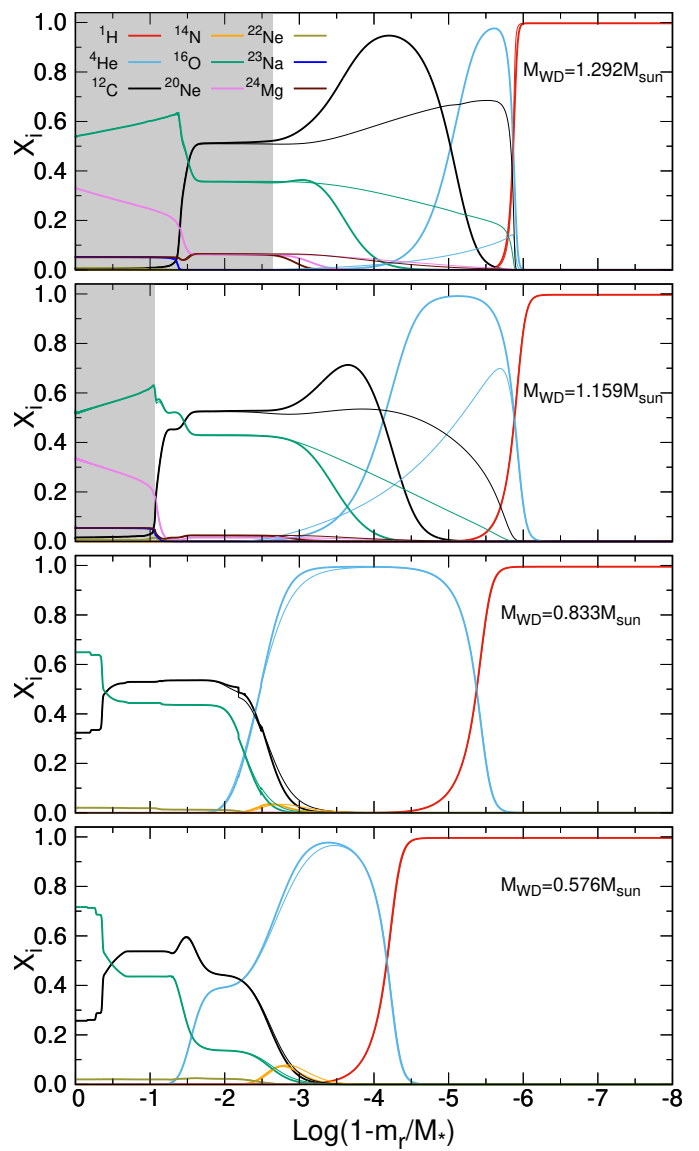}
        \caption{Same as Fig. \ref{perfiles_inicial.eps}, but at an effective 
        temperature of $T_{\rm eff} = 11\,000$ K, corresponding to the 
        domain of pulsating ZZ Ceti stars. Thick (thin) lines correspond
        to the case when Coulomb sedimentation of ions is considered (disregarded) in the
        diffusion equations. The gray area marks the domain of core crystallization.} 
        \label{perfiles_11000.eps}
\end{figure}

During  WD cooling,  element  diffusion processes  alter the  chemical
abundance  distribution in  the outer  layers  of all  of the  initial
models. In the case of ultra-massive WDs, phase separation of the core
chemical constituents  upon crystallization modifies the  shape of the
$^{16}$O and $^{22}$Ne  profiles. The resulting chemical profile  by the time
evolution   reached   the   ZZ   instability   strip   is   shown   in
Fig. \ref{perfiles_11000.eps} at $T_{\rm eff} = 11\,000$ K, which also
illustrates  the  impact  of  Coulomb sedimentation  on  the  chemical
profiles. As discussed, Coulomb  sedimentation is relevant for
massive WDs and for mixtures of  ions with equal $A/Z$.  In this case,
Coulomb    separation     drives    gravitational     settling    (see
Eqs. \ref{flux}-\ref{diff})  and, thus, ions  with larger $Z$  move to
deeper layers. We note that the H/He interface is not affected
by  Coulomb diffusion,  since in  this  case the  contribution due  to
gravity is dominant (Eq. \ref{grav}), and Coulomb diffusion represents
a minor  contribution to the  diffusion flux.  We also note  that for
ultra-massive WDs,  Coulomb
sedimentation prevents the strong inward diffusion of He toward the core,
leading to the formation of a pure-He buffer with the consequent 
appearance of two well separated chemical transition regions,
one located at the base and the other at the top of that buffer. 
The neglect of Coulomb diffusion in  ultra-massive WDs, on the contrary,
causes chemical diffusion, which  in this case is dominant (see
Eq. \ref{diff}), to virtually erode the initial pure He buffer, as
illustrated by the figure. Consider, for example, the case of the 
$1.292 M_\sun$ model (upper panel of Fig. \ref{perfiles_11000.eps}). 
When Coulomb diffusion is taken into account, there is the C/He interface 
located at $\log(1-m_r/M_{\star})\sim -5$, and the He/H interface, 
located at $\log(1-m_r/M_{\star})\sim -5.8$. On the contrary, 
for this model there is a single chemical transition region at 
$\log(1-m_r/M_{\star})\sim -5.8$ of $^{16}$O, $^{12}$C, $^{4}$He and $^{1}$H
when Coulomb diffusion is neglected. Finally, we note that 
when we consider Coulomb diffusion, there is an internal chemical 
transition (located in the fluid part of the star, at $\log(1-m_r/M_{\star})\sim -3$) 
due to the variation of chemical abundances of $^{24}$Mg, $^{20}$Ne, $^{16}$O, and $^{12}$C.  
This chemical transition is virtually absent in the case in which Coulomb 
diffusion is neglected. As we shall see, all these differences 
impact the theoretical  pulsational spectrum of these stars to
 some extent.
 
We also compare the resulting WD cooling times and find that the impact of
Coulomb diffusion on the cooling times is minor, expect for the  1.292\,$M_{\sun}$
WD sequence, which evolves about 6 $\%$ faster at low luminosities when
Coulomb diffusion is considered. For the
CO-core WD sequences, differences in the cooling times remain below  1 $\%$.

\begin{figure}
        \centering
        \includegraphics[width=1.0\columnwidth]{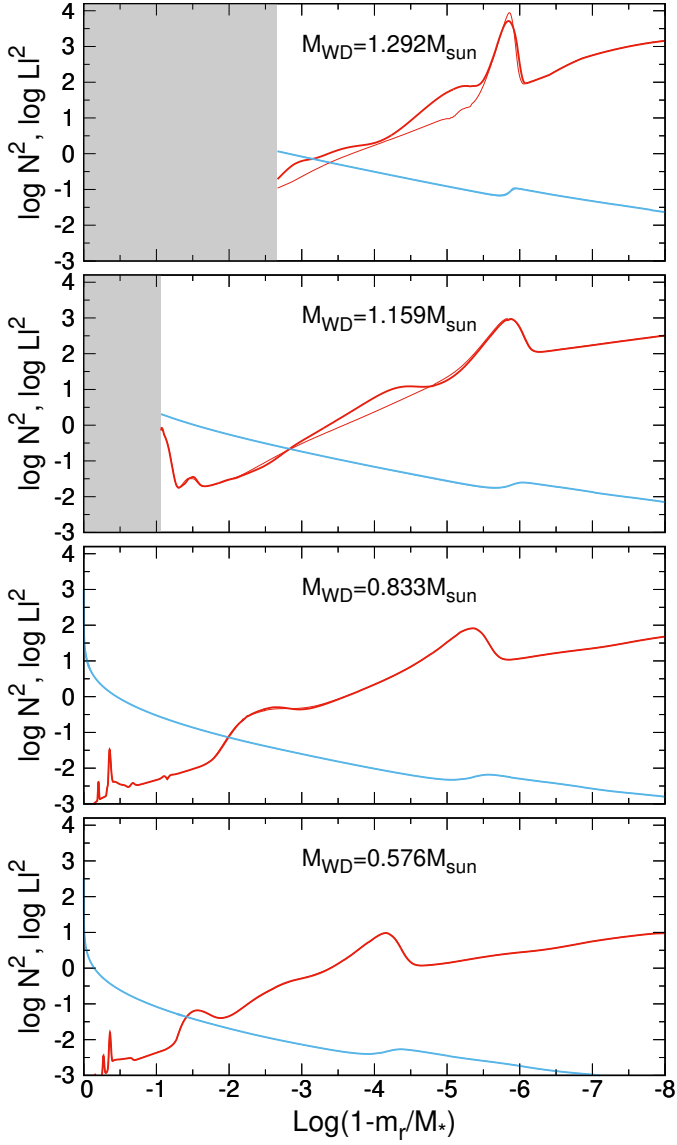}
        \caption{Logarithm of the squared Brunt-Va\"is\"al\"a and Lamb 
          frequencies (red and blue lines respectively) corresponding to the same models analyzed
          in Fig. \ref{perfiles_11000.eps} at  $T_{\rm eff}= 11\,000$ K.  Thick (thin) lines
          correspond  to the case when Coulomb sedimentation of ions is considered (disregarded)
          in the diffusion equations. The gray area marks the domain of core crystallization.
          The Lamb frequency corresponds to dipole ($\ell= 1$) modes.} 
        \label{bvf_zz.eps}
\end{figure}

%


\section{Pulsation results}
\label{pulsation_results}

\begin{figure}
        \centering
\includegraphics[width=1.0\columnwidth]{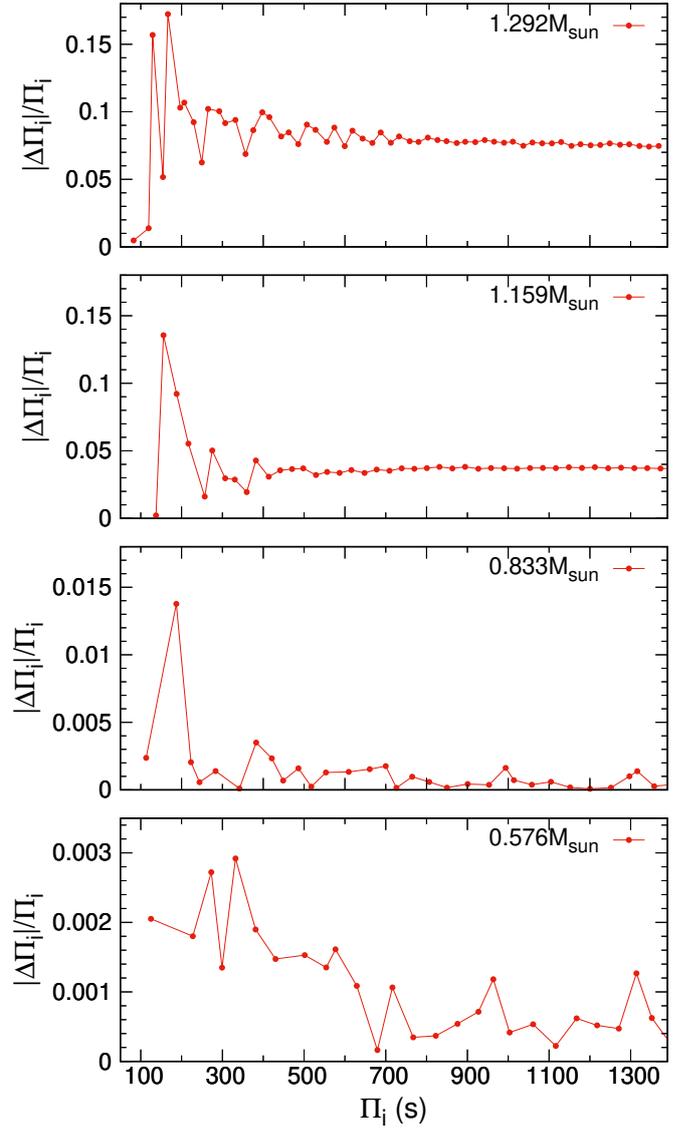}
        \caption{Relative period differences in terms of the periods
          of $\ell= 1$ pulsation $g$ modes resulting from including and disregarding 
          Coulomb separation
          in  the same models analyzed in Fig. \ref{perfiles_11000.eps} at  $T_{\rm eff}= 11\,000$ K.} 
        \label{deltaperiod.eps}
\end{figure}

In this section, we assess the impact of Coulomb separation of ions on
the  predicted  pulsation periods  of  our  WD  models.  We  begin  by
illustrating in  Fig.  \ref{bvf_zz.eps} the  changes in the run  of the
Brunt-V\"ais\"al\"a   and   Lamb   frequencies  induced   by   Coulomb
sedimentation for  all of our WD  models at $T_{\rm eff}=  11\,000$ K.
In particular, the 1.16 and 1.29 $M_{\odot}$ models start to crystallize
before reaching the blue edge of the ZZ Ceti instability strip. In these 
cases, the gray area  marks the domain of core  crystallization. 
Any chemical interface located within  the crystallized region has  
no relevance to the pulsation  properties of  the $g$ modes, 
that is why we have hidden them under the gray area in the plots.   
The Brunt-V\"ais\"al\"a frequency reflects any change in  chemical 
composition in the interior
of the  model.  Indeed,  there
exist dominant bumps  in the run of  the Brunt-V\"ais\"al\"a frequency
associated to  the various  chemical transition  regions, particularly
the  H/He  interface,  which  produces  the  most  dominant
feature   in   the   Brunt-V\"ais\"al\"a  frequency. We note that,
in the case of the $1.292 M_\sun$  model computed with Coulomb diffusion,
there exist two bumps, one of them due to the chemical transition region 
of C and He, and the other one associated to the He/H chemical interface  
(see Fig. \ref{perfiles_11000.eps}). At variance with this, in the case 
in which Coulomb diffusion is neglected, there is a single bump in the Brunt-V\"ais\"al\"a 
frequency which is due to the chemical interface of O, C, He and H.
On the other hand, there is an additional internal bump in the
Brunt-V\"ais\"al\"a frequency at $\log(1-m_r/M_{\star})\sim -3$ 
that is due to the multiple chemical transition 
region of $^{24}$Mg, $^{20}$Ne, $^{16}$O, and $^{12}$C (see also Fig. \ref{perfiles_11000.eps}), which is absent in the case 
in which Coulomb diffusion is neglected. This bump is located in the 
fluid part of the star, that is, within the mode propagation zone, 
and therefore it has a non-zero impact on the mode trapping properties of 
$g$ modes. 

The shape of the Brunt-V\"ais\"al\"a  frequency has a strong impact on
the $g$-mode period spectrum and mode-trapping properties of pulsating
WDs.  In  view of  the  above  discussion about the
Brunt-V\"ais\"al\"a frequency, we expect that Coulomb diffusion of ions alters
markedly the $g$-mode pulsation periods of massive ZZ Ceti stars. This
is borne out by Fig.  \ref{deltaperiod.eps}, that displays the relative
period differences in terms of the  periods of $\ell= 1$ pulsation $g$
modes resulting from including  and disregarding Coulomb separation in
our  WD   models at $T_{\rm  eff}=   11\,000$  K. The differences are 
defined as the periods calculated with Coulomb diffusion minus 
the periods neglecting this effect, divided by the periods computed 
with Coulomb diffusion. This relative difference is plotted in  
terms of the periods calculated with Coulomb diffusion. We note  that   for  the
ultra-massive WDs ($M_{\star} > 1.16 M_{\odot}$), Coulomb diffusion  
yields significant changes in the $g$-mode pulsation periods by as much  
as $\sim 15 \%$, which is given for the lowest radial-order modes 
($\Pi \lesssim 600$ s).   This is a relevant effect   and    should   
be    taken   into   account    in   detailed
asteroseismological analyses  of these  pulsating stars. On  the other
hand, the  impact on less  massive WDs  is much less  noticeable, with
changes in the pulsation periods  no larger than $0.3\%$.  
Such period changes remain still above the typical uncertainties of the 
observed periods ($\lesssim 0.01 \%$) but  well  below the period  changes 
that result from  uncertainties in convective boundary mixing and  nuclear 
reactions during WD progenitor evolution \citep[see][]{2017A&A...599A..21D}.

\section{Summary and conclusions}
\label{conclusions}

Motivated by the result of  \cite{2013PhRvL.111p1101B} who suggest that the 
redistribution of ions due to Coulomb  separation could affect  the thermal 
evolution of  WDs and their  pulsational  properties, we have undertaken the present
investigation to precisely assess the impact of such Coulomb separation on  the WD evolution as well as on their  chemical profiles, the Brunt-V\"ais\"al\"a frequency, and their pulsational periods at the  ZZ Ceti instability strip. To this end, we have followed the full evolution of white dwarf  models in the  range 0.5-1.3 $ M_\sun$ derived from their progenitor  history on the basis of a time-dependent   element  diffusion   scheme  that   incorporates  the effect of gravitational  settling  of ions due  to Coulomb  interactions  at  high densities.

We find that Coulomb sedimentation profoundly alters the chemical profiles of ultra-massive white dwarf along their evolution, preventing the strong inward diffusion of He toward the core, and thus leading 
to the formation of a pure-He buffer with two separate chemical interfaces, at variance with 
what happens when Coulomb diffusion is neglected.
These changes in the
inner chemical distribution barely affect the cooling properties of the WDs.
However, the impact on the pulsation periods is quite different. In fact, large changes in the $g$-modes pulsation periods as high as $15 \%$ are
expected for ultra-massive ZZ Ceti stars. For less massive ZZ Ceti stars, the impact
of Coulomb separation is much less noticeable,  inflicting period changes that are below the period changes that result from uncertainties in progenitor
evolution, albeit larger than typical uncertainties of observed periods.  
The process of Coulomb diffusion of ions profoundly affects the diffusion flux in massive white dwarfs, driving  the  gravitational settling of ions with the same $A/Z$ (mass to charge number). In this study, we have quantified the magnitude of the impact of such effect on the period spectrum of ZZ Ceti stars. The changes in the pulsational periods  are sufficiently important  that
they should  not be neglected  in   detailed asteroseismological analyses  of massive ZZ Ceti stars.

\begin{acknowledgements}
 
Part of  this work was  supported by PICT-2017-0884 from ANPCyT, PIP
112-200801-00940 grant from CONICET, grant G149 from University of La Plata.
This  research has  made use of  NASA Astrophysics Data System.
\end{acknowledgements}

\bibliographystyle{aa}
\bibliography{ultramassiveCO}



\end{document}